\documentclass[10pt]{iopart}

\usepackage{graphicx}
\usepackage{dcolumn}
\usepackage{bm}
\usepackage{color}

\begin{document}

\title[ECH Heliotron J]{Analysis of the ECH effect on the EPM/AEs stability in Heliotron J plasma using a Landau closure model}


\author{J. Varela}
\ead{jvrodrig@fis.uc3m.es}
\address{Universidad Carlos III de Madrid, 28911 Leganes, Madrid, Spain}
\author{K. Nagasaki}
\address{Institute of Advanced Energy, Kyoto University, Uji, Kyoto, Japan}
\author{S. Kobayashi}
\address{Institute of Advanced Energy, Kyoto University, Uji, Kyoto, Japan}
\author{K. Nagaoka}
\address{National Institute for Fusion Science, National Institute of Natural Science, Toki, 509-5292, Japan}
\author{P. Adulsiriswad}
\address{National Institute for Fusion Science, National Institute of Natural Science, Toki, 509-5292, Japan}
\author{A. Cappa}
\address{Laboratorio Nacional de Fusion CIEMAT, Madrid, Spain}
\author{S. Yamamoto}
\address{National Institutes for Quantum and Radiological Science and Technology, Naka, Ibaraki 311-0193, Japan}
\author{K.Y. Watanabe}
\address{National Institute for Fusion Science, National Institute of Natural Science, Toki, 509-5292, Japan}
\author{D. A. Spong}
\address{Oak Ridge National Laboratory, Oak Ridge, Tennessee 37831-8071, USA}
\author{L. Garcia}
\address{Universidad Carlos III de Madrid, 28911 Leganes, Madrid, Spain}
\author{Y. Ghai}
\address{Oak Ridge National Laboratory, Oak Ridge, Tennessee 37831-8071, USA}
\author{J. Ortiz}
\address{Universidad Carlos III de Madrid, 28911 Leganes, Madrid, Spain}

\date{\today}

\begin{abstract}
The aim of the present study is to analyze the effect of the electron cyclotron heating (ECH) on the linear stability of Alfven Eigenmodes (AE) and energetic particle modes (EPM) triggered by energetic ions in Heliotron J plasma. The analysis is performed using the FAR3d code that solves a reduced MHD model to describe the thermal plasma coupled with a gyrofluid model for the energetic particles (EP) species. The simulations reproduce the AE/EPM stability trends observed in the experiments as the electron temperature ($T_{e}$) increases, modifying the thermal plasma $\beta$, EP $\beta$ and EP slowing down time. Particularly, the $n/m=1/2$ EPM and $2/4$ Global AE (GAE) are stabilized in the low bumpiness (LB) configuration due to an enhancement of the continuum, Finite Larmor radius (FLR) and e-i Landau damping effects as the thermal $\beta$ increases. On the other hand, a larger ECH injection power cannot stabilize the AE/EPM in Medium (MB) and High bumpiness (HB) configurations because the damping effects are weaker compared to the LB case, unable to balance the further destabilization induced by an enhanced EP resonance as the EP slowing down time and EP $\beta$ increases with $T_{e}$. 

\end{abstract}

%
%
%
%
%


\vspace{2pc}

\maketitle

\ioptwocol

\section{Introduction \label{sec:introduction}}

There are several techniques dedicated to improve the Alf\'enic stability of nuclear fusion plasma. For example, the operational regime of energetic particle (EP) sources can be optimized to reduce the plasma perturbation \cite{1,2,3,4,5,6,7,8}. Other example is the local modification of the magnetic trap by non inductive currents generated by the electron cyclotron current drive (ECCD) \cite{9,10,11} or the neutral beam current drive (NBCD) \cite{12,13,14}. Another option is increasing the plasma temperature using electron cyclotron heating (ECH) to modify locally the EP slowing-down distribution function and the damping effects as the thermal $\beta$ grows \cite{15,16,17,18,19,20}. In addition, recent analysis are dedicated to study the co-existence of different EP populations in reactor relevant plasma, that is to say, plasma with fusion born alpha particles and EP generated by neutral beam injectors (NBI) at the same time \cite{21,22,23}.

The injection of electron cyclotron waves (ECW) \cite{24,25} generates non inductive currents and heats the plasma \cite{18,26,27}, leading to the stabilization or further enhancement of the Alfven Eigenmodes (AE) or energetic particle modes (EPM) depending on the injector power and configuration \cite{19,28}.

AE / EPM can be triggered if there is a resonance between the EP drift, bounce or transit frequencies and the AE / EPM frequency \cite{29}. Unstable AE / EPM cause losses of EP before thermalization, enhancing the transport of fusion produced alpha particles and the EP generated by NBI, ECW and ion cyclotron wave (ICW) \cite{30,31,32}. Consequently, the heating efficiency of the nuclear device decreases \cite{33,34,35}.

Heliotron J is a medium-sized helical device with four toroidal magnetic field periods \cite{36,37}. The coil system is composed of an $L/M = 1/4$ helical coil as well as two kind of toroidal coils called inner and outer vertical coils. The magnetic configuration can be controlled by varying the current ratios in each coil. Three configurations are chosen in the paper, the high-, medium-, and low-bumpiness configurations with fixed toroidicity, helicity, rotational transform and plasma volume, where bumpiness means the toroidal variation of the magnetic field strength \cite{38}.

Plasmas are produced and heated by second-harmonic X-mode $70$ GHz ECH and NBI. In addition, ECH is applied with a $70$ GHz 2nd X-mode configuration for a total injection power of $0.4$ MW \cite{39}. Two NBI tangential hydrogen beam lines, BL1 and BL2, are used, both of which have a maximum acceleration voltage of $30$ keV and a maximum power of $0.8$ MW \cite{40}. The injected ECH power is as high as $0.3$ MW, and the NBI power is as high as $1.3$ MW in total in the experiment reported here.

The experiments show the application of ECH can lead to the stabilization or further destabilization of the AE/EPM triggered by energetic ions in NBI heated plasma depending on the ECH injection power and the Heliotron J magnetic configuration \cite{18,19}. The apparent complexity of the ECH effect on the plasma MHD stability can be explained by the combined variation of the EP $\beta$, EP resonance properties, plasma resistivity and continuum + FLR + e-i Landau dampings effects as the electron temperature increases. 

Present study is dedicated to analyze the AE / EPM stability in Heliotron J discharges for different ECH injection powers and magnetic configurations, identifying the MHD stability trends with respect to the EP $\beta$, EP slowing down time and thermal $\beta$ (including the effect of the plasma resistivity as well as the continuum, FLR and e-i Landau damping effects). The study is performed using the FAR3d gyro-fluid code \cite{41,42,43,44} that solves the reduced linear resistive MHD equations coupled with the EP density and parallel velocity equations \cite{45,46,47}. The FAR3d code includes the linear wave-particle resonance to reproduce the Landau damping/growth by Landau closure relations, analyzing the evolution of six field variables in a three dimensional equilibria generated by the VMEC code \cite{48}. 

This paper is organized as follows. The numerical scheme and equilibrium properties are described in section \ref{sec:model}. The linear stability of the AE/EPM is studied for the low, medium and high bumpiness Heliotron J configurations in section \ref{sec:lin}. Next, the conclusions of the study are shown in section \ref{sec:conclusions}.

\section{Numerical scheme \label{sec:model}}

The numerical model solves a set of reduced MHD equations retaining the toroidal angle dependency in a three-dimensional VMEC equilibrium \cite{48,49}. The effect of the EP perturbation is included by moments of the gyro-kinetic equation distribution function: the EP density ($n_{f}$) and the EP velocity parallel to the magnetic field lines ($v_{||f}$). Landau closure coefficients are required to truncate the number of gyro-kinetic equation moments included in the model. The closure is obtained from gyro-kinetic simulations, matching the analytic TAE growth rates of the two-pole approximation of the plasma dispersion function, leading to a Lorentzian energy distribution function for the EP. The 2-moment gyrofluid model used here is based on a Lorentzian distribution function that is matched to a Maxwellian or to a slowing-down distribution by choosing an equivalent average energy. The EP distribution in the simulations has the same second moment, the effective EP temperature, as that of the equivalent slowing down distribution. It should be noted that a single EP Maxwellian distribution cannot reproduce the same resonance as a slowing down distribution, because the gradient of the phase space distribution determines the drive of the AE modes. Nevertheless, a set of Maxwellian distribution functions can be used to approximate the resonances triggered by a slowing down distribution function. For this reason, we perform parametric analysis with respect to the EP energy and $\beta$. Please see the references \cite{50,51} for further details of the model equations and numerical scheme.

FAR3d code participated in benchmarking studies validating the model results with respect to gyro-kinetic and hybrid codes \cite{52}. Previous studies performed using FAR3d show a reasonable agreement with the experimental data, for example reproducing the AE stability in Heliotron J \cite{53,54}, LHD \cite{55,56}, TJ-II \cite{8,44,57} and DIII-D \cite{58,59} plasma. Likewise, nonlinear simulations reproduced the sawtooth-like events, internal collapse and EIC burst observed in LHD \cite{60,61,62,63,64} and the AE saturation in DIII-D \cite{65}. Also, FAR3d simulations of Heliotron J plasma show similar results compared to MEGA code modeling \cite{66,67}. In particular, the interaction between the high-velocity EPs that transit the core region and the peripheral $n/m=1/2$ EPM using free boundary conditions by MEGA code, and the role of the EP profiles near the plasma periphery performing parametric studies by FAR3d, identifying how the EP driving rate of $1/2$ EPM and the $2/4$ GAE change through variations of the mode spatial profile.

\subsection{Equilibrium properties}

\begin{figure}[h!]
\centering
\includegraphics[width=0.45\textwidth]{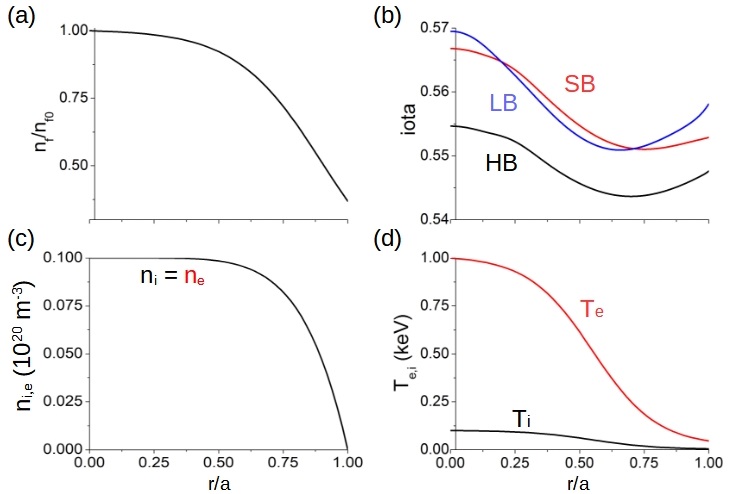}
\caption{(a) Normalized EP density profile, (b) iota profile for each Heliotron J magnetic configuration, (c) thermal plasma density, (d) thermal plasma temperature.}\label{FIG:1}
\end{figure}

Three equilibria of Heliotron J discharges with low (LB), medium (MB) and high (HB) bumpiness are calculated using the VMEC code \cite{37}. The fuelling gas is hydrogen, the major radius $1.2$ m and the magnetic field intensity at the magnetic axis is $1.25$ T. The ECH injection in the experiments leads to an enhancement of the electron temperature in the magnetic axis from $0.5$ to $1$ keV as the ECH power increases from $100$ to $300$ kW. The simulations analyze the plasma stability if the electron temperature at the magnetic axis increases from $0.5$ to $2$ keV. Figure~\ref{FIG:1} shows the model profiles for the thermal plasma and EP.

The pressure profile shape is assumed the same in the three magnetic configurations because the experimental data shows few differences in the thermal plasma profiles. However, the peak value of the temperature increases with ECH heating. This results in an increment of the thermal pressure as the electron temperature increases; it is reproduced by scaling up the thermal pressure profile in the simulations, that is to say, the radial shape of the thermal pressure is fixed, but the absolute values increase. This approximation is also valid because the thermal $\beta$ of the experiment is small ($\approx 0.2 \%$). For simplicity, no radial dependency of the EP energy is assumed. The range of nominal EP energies analyzed goes from $6$ to $25$ keV ($v_{th,f0} / V_{A0} = 0.15$ - $0.31$). Here, $v_{th,f0}$ is the thermalized EP velocity and $V_{A0}$ the Alfven velocity. The EP $\beta$ values studied ($\beta_{f}$) go from $0.0005$ to $0.01$.

\subsection{Simulation parameters}

The dynamic and equilibrium toroidal ($n$) and poloidal ($m$) modes included in the simulations are listed in the table~\ref{Table:1}. The simulations only include the toroidal families $n=1$ to $2$ in the LB configuration, mode selection extended to $n=6$ in MB and HB configurations because such discharges show a stronger AE activity and higher toroidal mode families could be unstable. The poloidal modes in the simulations are chosen to include all the resonant modes between the magnetic axis and the plasma periphery. The number of radial grid points is $1000$.  

\begin{table}[h]
\centering
\begin{tabular}{c | c }
\hline
$n$ & $m$\\ \hline
$0$ & $[0 , 14]$ \\
$1$ & $[1 , 3]$ \\
$2$ & $[2 , 6]$ \\
$3$ & $[3 , 9]$ \\
$4$ & $[4 , 12]$ \\
$5$ & $[5 , 15]$ \\
$6$ & $[6 , 18]$ \\
\end{tabular}
\caption{Dynamic and equilibrium toroidal (n) and poloidal (m) modes in the simulations.} \label{Table:1}
\end{table}

The dynamic variables must include both mode parities because the moments of the gyro-kinetic equation breaks the MHD symmetry (for more details please see \cite{55}). The magnetic Lundquist number ranges between $S=10^6$ - $10^7$ and the normalized Larmor radius of the thermal ions (with respect to the minor radius) goes from $0.0091$ to $0.0183$ depending on the electron temperature. The normalized EP Larmor radius is $0.05$.

Eigenfunctions (f) in FAR3d code are represented in terms of sine and cosine components, using real variables:
\begin{eqnarray} 
f(\rho,\theta, \zeta, t) = \sum_{m,n} f^{s}_{mn}(\rho, t) sin(m \theta + n \zeta) \nonumber\\
+ \sum_{m,n} f^{c}_{mn}(\rho, t) cos(m \theta + n \zeta)
\end{eqnarray}
In the following, the cosine component of the eigenfunction is indicated by positive mode numbers and the sine components by negative mode numbers.

\section{ECH injection effect on the AE/EPM stability in Heliotron J discharges \label{sec:lin}}

The analysis consists in a set of simulations reproducing the effect of a stronger ECH injection power (higher electron temperature) on the EP slowing down time, EP $\beta$ and thermal plasma $\beta$. That way, the dominant trends of the AE/EPM stability can be identified with respect to the variation of different plasma parameters as the electron temperature increases. The experimental observations show clear differences regarding the impact of the ECH injection power on the AE/EPM stability for each magnetic configuration. In the following, the AE/EPM stability is analyzed for each magnetic configuration individually.

\subsection{Low bumpiness configuration}

\begin{figure}[h!]
\centering
\includegraphics[width=0.45\textwidth]{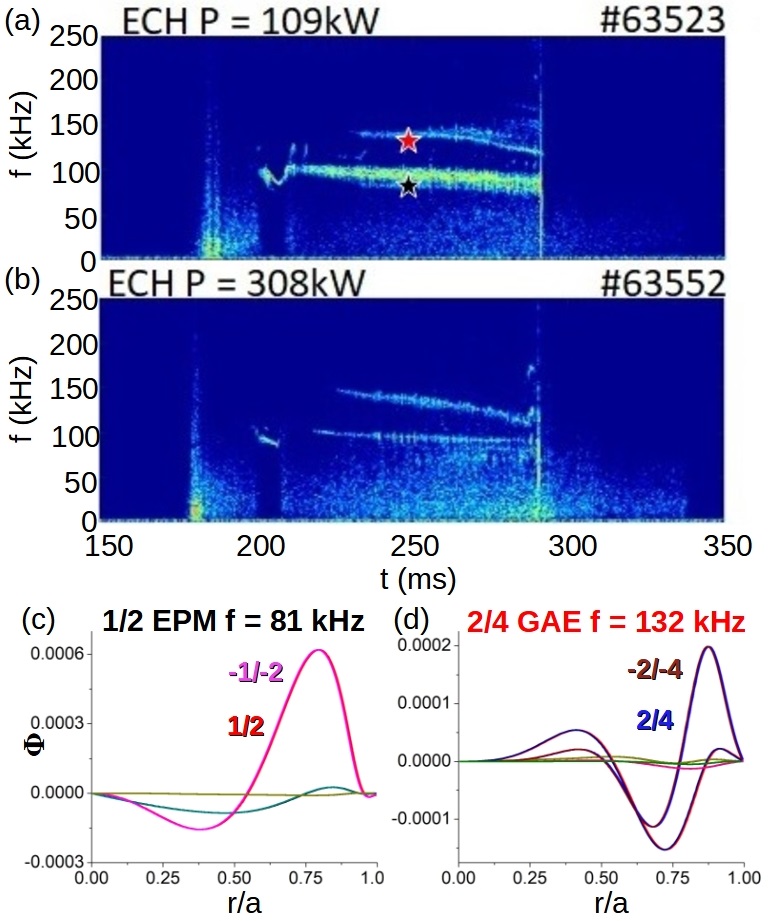}
\caption{Magnetic spectrogram of discharges with an ECH injection power of (a) $100$ kW and (b) $300$ kW. The colored stars indicate the frequency range of the modes calculated by FAR3d (black $n=1$ and red $n=2$ modes). Mode eigenfunction calculated by FAR3d: (c) $1/2$ EPM and (d) $2/4$ GAE.}\label{FIG:2}
\end{figure}

The LB configuration shows the lowest AE/EPM activity, stabilized above a given threshold of the ECH injection power \cite{19}. Figure~\ref{FIG:2} shows the magnetic spectrogram of two discharges with different ECH injection power. The AE/EPM are almost stabilized if the ECH power increases from $100$ kW (panel a) to $300$ kW (panel b). The colored stars indicate the frequency range of the dominant modes calculated by the FAR3d code (for an EP $\beta = 0.003$, $T_{f} = 14$ keV and $T_{e} = 0.5$ keV). Panels c and d show the modes eigenfunction. FAR3d simulations reproduces the same dominant mode, frequency range and radial location with respect to the experiment observations \cite{19,54}. The modes are the $n/m=1/2$ EPM with $f = 81$ kHz (panel c) and the $2/4$ GAE with $132$ kHz (panel d), both destabilized in the plasma periphery. 

\begin{figure}[h!]
\centering
\includegraphics[width=0.45\textwidth]{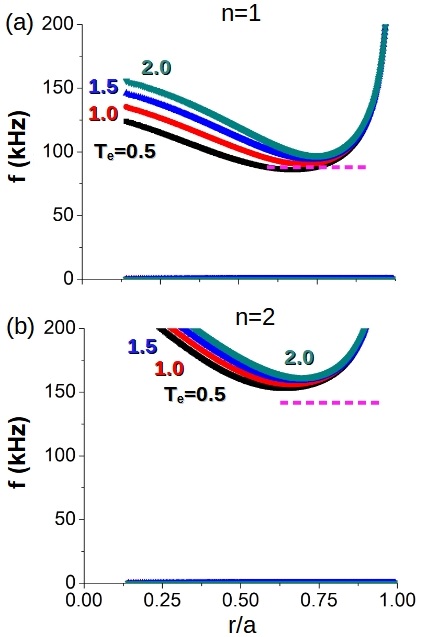}
\caption{Alfv\'en gap structure for different electron temperatures in LB configuration for (a) $n=1$ and (b) $n=2$ toroidal families. $T_{e} = 0.5$ keV (black line), $T_{e} = 1.0$ keV (red line), $T_{e} = 1.5$ keV (blue line) and $T_{e} = 2.0$ keV (cyan line). The dashed pink horizontal lines indicate the frequency range and eigenfunction width of the AE/EPM if EP $\beta = 0.01$.}\label{FIG:3}
\end{figure}

The increment of the electron temperature as the ECH injection power enhances modifies the Alfv\'en gap structure. Figure~\ref{FIG:3} shows the evolution of the Alfven gaps as the electron temperature at the magnetic axis increases from $0.5$ to $2$ keV. The Alfven gaps are calculated using the code STELLGAP including the effect of the sound wave coupling \cite{68}. A higher electron temperature leads to a frequency up-shift of the Alfv\'en gaps, particularly in the inner plasma region, as well as an outward displacement of the gaps frequency minima. Consequently, the stabilizing effect of the continuum damping on the AE/EPM may change. The horizontal pink dashed lines indicate the frequency range and radial location of the AE/EPM obtained in the simulations with $T_{e} = 1$ keV and EP $\beta = 0.01$. The $1/2$ EPM is destabilized inside the continuum and the $2/4$ GAE nearby the local minima of the gap frequency, consistent with the mode identification in figure~\ref{FIG:2}.

\begin{figure}[h!]
\centering
\includegraphics[width=0.45\textwidth]{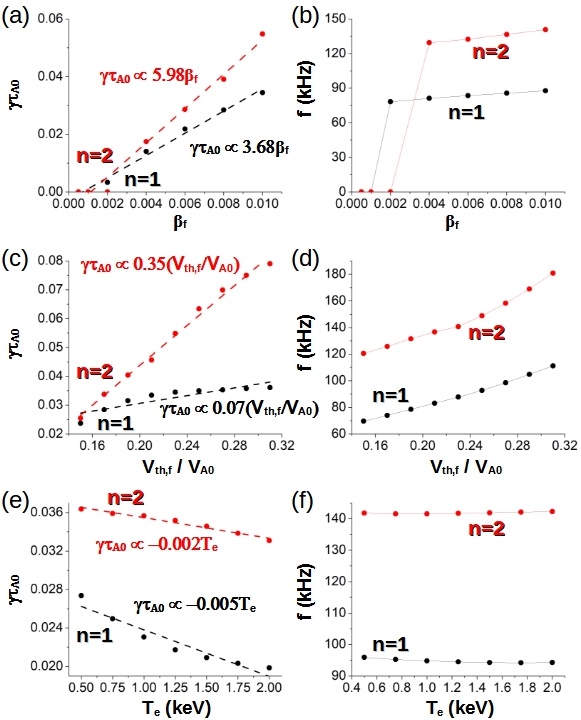}
\caption{AE/EPM stability trends of the dominant modes in the LB configuration. (a) Growth rate and (b) frequency of the $1/2$ EPM and $2/4$ GAE for different EP $\beta$ values. (c) Growth rate and (d) frequency of the $1/2$ EPM and $2/4$ GAE for different $v_{th,f0} / V_{A0}$ ratios. (e) Growth rate and (f) frequency of the $1/2$ EPM and $2/4$ GAE for different electron temperatures. The dashed lines indicate the linear regressions: $\gamma \tau_{A0} = A\beta_{f}$, $\gamma \tau_{A0} = B(v_{th,f0} / V_{A0})$ and $\gamma \tau_{A0} = CT_{e}$.}\label{FIG:4}
\end{figure}

Next, the effect of the ECH injection power on the AE/EPM stability with respect to the EP $\beta$, EP slowing down time and thermal $\beta$ is analyzed. A higher $T_{e}$ leads to an increase of the EP $\beta$, EP slowing down time and thermal $\beta$. Figure~\ref{FIG:4} panels a and b show the EP $\beta$ threshold is $0.002$ for the $1/2$ EPM and $0.004$ for the $2/4$ GAE (fixed $v_{th,f0} / V_{A0} = 0.23$ and $T_{e} = 1$ keV). A larger EP population causes a stronger destabilization of the modes, reason why the simulation growth rate increases with the EP $\beta$. The linear regression of the growth rate with respect to the EP $\beta$ shows a slope of $3.68$ for the $1/2$ EPM and $5.98$ for the $2/4$ GAE. Increases in the EP slowing down time are evaluated as increases of the $v_{th,f0} / V_{A0}$ ratio, that is to say, via modification of the EP resonance as the EP energy increases. Figure~\ref{FIG:4} panels c and d show a larger growth rate and frequency as the $v_{th,f0} / V_{A0}$ ratio increases (simulation with fixed EP $\beta = 0.01$ and $T_{e} = 1$ keV). The linear regression of the growth rate with respect to the velocity ratio indicates a slope of $0.07$ for the $1/2$ EPM and $0.35$ for the $2/4$ GAE. Figure~\ref{FIG:4} panels e and f show the growth rate and frequency as the thermal $\beta$ increases (fixed $v_{th,f0} / V_{A0} = 0.23$ and $\beta = 0.01$), including the decrease of the plasma resistivity as well as the variation of the continuum, thermal ion FLR, EP FLR and electron-ion Landau dampings. There is a decrease of the growth rate as the thermal $\beta$ increases although the frequency is weakly affected. The slope of the linear regression is $-0.005$ for the $1/2$ EPM and $-0.002$ for the $2/4$ GAE. Following the trends calculated, the growth rate decrease by $15 \%$ for the $1/2$ EPM and by $2 \%$ for the $2/4$ GAE as the thermal $\beta$ increases comparing simulations with $T_{e} = 0.5$ and $1.0$ keV. That means, the stabilizing trend linked to the increment of the thermal $\beta$ can be only compensated if the EP $\beta$ or the velocity ratio increase by $ \Delta \beta_{f} = 0.00077$ ($7.1 \%$) or $\Delta (v_{th,f0} / V_{A0}) = 0.034$ ($16 \%$) for the $1/2$ EPM, as well as by $ \Delta \beta_{f} = 0.0006$ ($6 \%$) or $\Delta (v_{th,f0} / V_{A0}) = 0.07$ ($23 \%$) for the $2/4$ GAE. Such increments are too large for a $T_{e}$ difference of $\Delta T_{e} = 0.5$ keV. Consequently, the stabilizing effect linked to the increment of the thermal $\beta$ is dominant, particularly for the $n=1$ EPM.

\begin{figure}[h!]
\centering
\includegraphics[width=0.45\textwidth]{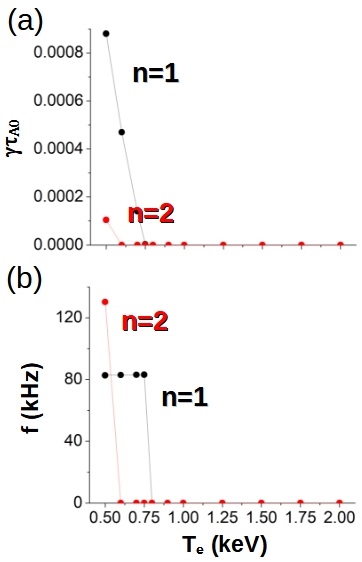}
\caption{(a) Growth rate and (b) frequency of the $1/2$ EPM and $2/4$ GAE for different electron temperature values. EP $\beta = 0.003$.}\label{FIG:5}
\end{figure}

Identified the dominant stability trends, new simulations are performed using a set of parameters as close as possible to the experimental conditions for different ECH injection powers. The simulations have an EP $\beta = 0.003$, including the EP / thermal ion FLR and electron-ion Landau damping effects. Figure~\ref{FIG:5} shows the AE/EPM growth rate and frequency as the electron temperature increases. The simulations indicate the stabilization of the $1/2$ EPM and $2/4$ GAE as the electron temperature grows from $0.5$ to $1$ keV, consistent with the experimental observations.

Summarizing, the AE/EPM are stabilized if the ECH injection power is $300$ kW because the effects of the continuum, FLR and electron-ion Landau damping are dominant for a plasma with $T_{e} = 1$ keV. That means, the enhancement of the damping effects as $T_{e}$ increases compensate the further destabilization caused by a larger slowing down time and $\beta$ of the EP. 

\subsection{Medium bumpiness configuration}

\begin{figure}[h!]
\centering
\includegraphics[width=0.5\textwidth]{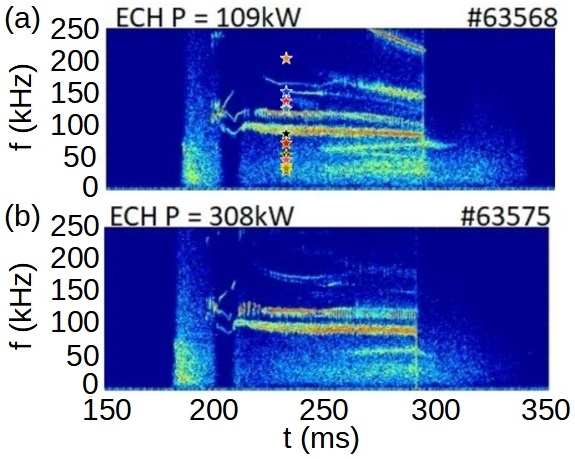}
\caption{Magnetic spectrogram of discharges with an ECH injection power of (a) $100$ kW and (b) $300$ kW. The colored stars indicate the frequency range of the dominant and sub-dominant modes calculated by FAR3d (black $n=1$, red $n=2$, $n=3$ blue, $n=4$ cyan, $n=5$ pink and $n=6$ yellow).}\label{FIG:6}
\end{figure}

Figure~\ref{FIG:6} shows the magnetic spectrogram of two discharges with an ECH injection power of $100$ kW (panel a) and $300$ kW (panel b) for a medium bumpiness configuration. The increment of the ECH injection power weakly affects the two strongest alfvenic instabilities observed in the frequency range between $100$ and $135$ kHz. On the other hand, the alfvenic instabilities below $100$ kHz and above $150$ kHz are fully or partially stabilized. The colored stars show the frequency range of the dominant and sub-dominant modes calculated by FAR3d if EP $\beta = 0.003$, $T_{f} = 14$ keV and $T_{e} = 0.5$ keV. The identification of the dominant and sub-dominant modes is done in the Appendix 1. The analysis indicate the modes with the largest growth rate are the $1/2$ GAE with $85$ kHz, $2/4$ GAE with $134$ kHz, $3/5$ GAE with $151$ kHz and $4/7$ EPM with $135$ kHz.

\begin{figure}[h!]
\centering
\includegraphics[width=0.45\textwidth]{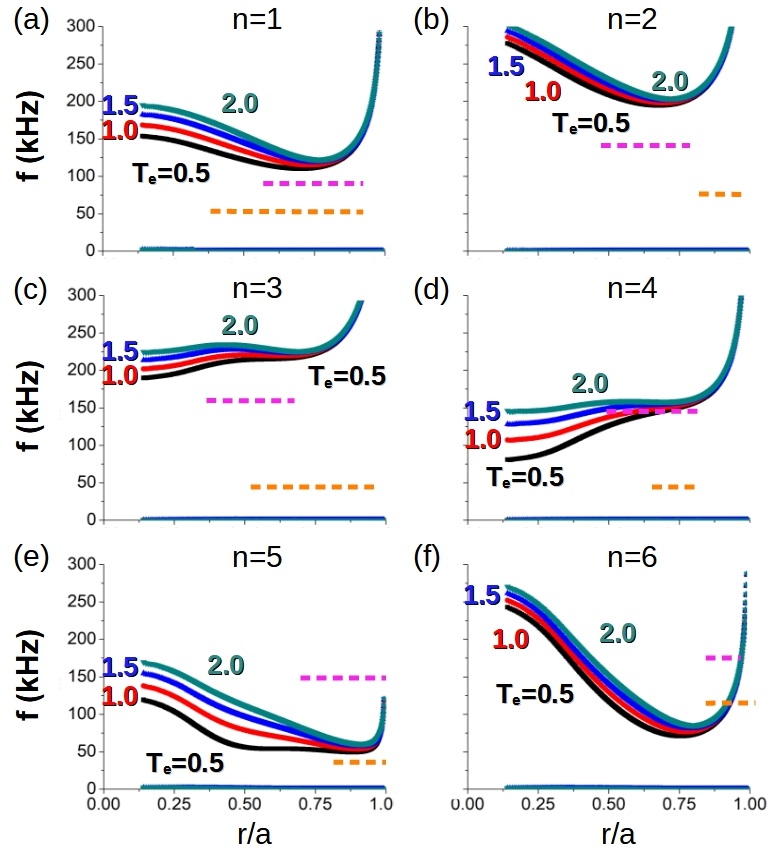}
\caption{Alfv\'en gap structure for different electron temperatures in MB configuration for the (a) $n=1$, (b) $n=2$, (c) $n=3$, (d) $n=4$, (e) $n=5$ and (f) $n=6$ toroidal mode families. $T_{e} = 0.5$ keV (black line), $T_{e} = 1.0$ keV (red line), $T_{e} = 1.5$ keV (blue line) and $T_{e} = 2.0$ keV (cyan line). The dashed pink (orange) horizontal lines indicate the frequency range and eigenfunction width of the dominant (sub-dominant) AE/EPM if EP $\beta = 0.01$.}\label{FIG:7}
\end{figure}

Figure~\ref{FIG:7} shows the evolution of the Alfven gaps as the electron temperature t the magnetic axis increases from $0.5$ to $2$ keV. The number of toroidal mode families analyzed goes up to $n=6$. Again, a higher electron temperature leads to a frequency up-shift of the Alfv\'en gaps in the inner-middle plasma region. The horizontal pink (orange) dashed lines indicate the frequency range and radial location of the dominant (sub-dominant) AE/EPM obtained in FAR3d simulations with $T_{e} = 1$ keV and EP $\beta = 0.01$. The unstable modes are located in the middle-outer plasma region showing a large frequency spreading, from $35$ to $175$ kHz.

\begin{figure}[h!]
\centering
\includegraphics[width=0.45\textwidth]{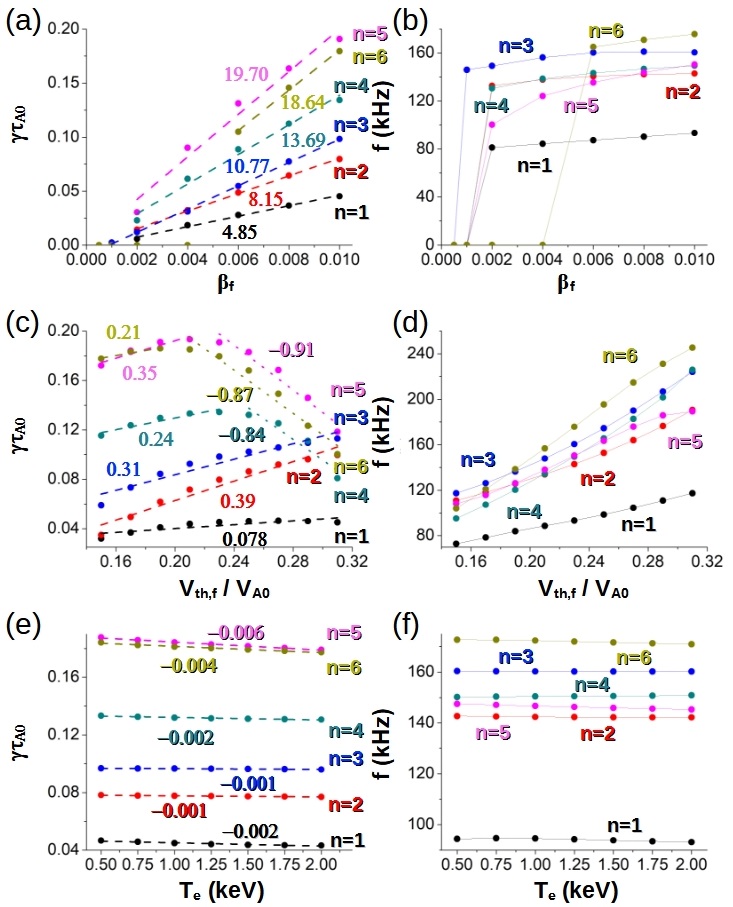}
\caption{AE/EPM stability trends of the dominant modes in the MB configuration. (a) Growth rate and (b) frequency of the $n=1$ to $6$ modes for different EP $\beta$ values. (c) Growth rate and (d) frequency of the $n=1$ to $6$ modes for different $v_{th,f0} / V_{A0}$ ratios. (e) Growth rate and (f) frequency of the $n=1$ to $6$ modes for different electron temperatures. The dashed lines indicate the result of the linear regressions: $\gamma \tau_{A0} = A\beta_{f}$, $\gamma \tau_{A0} = B(V_{th,f0} / V_{A0})$ and $\gamma \tau_{A0} = CT_{e}$.}\label{FIG:8}
\end{figure}

\begin{figure}[h!]
\centering
\includegraphics[width=0.45\textwidth]{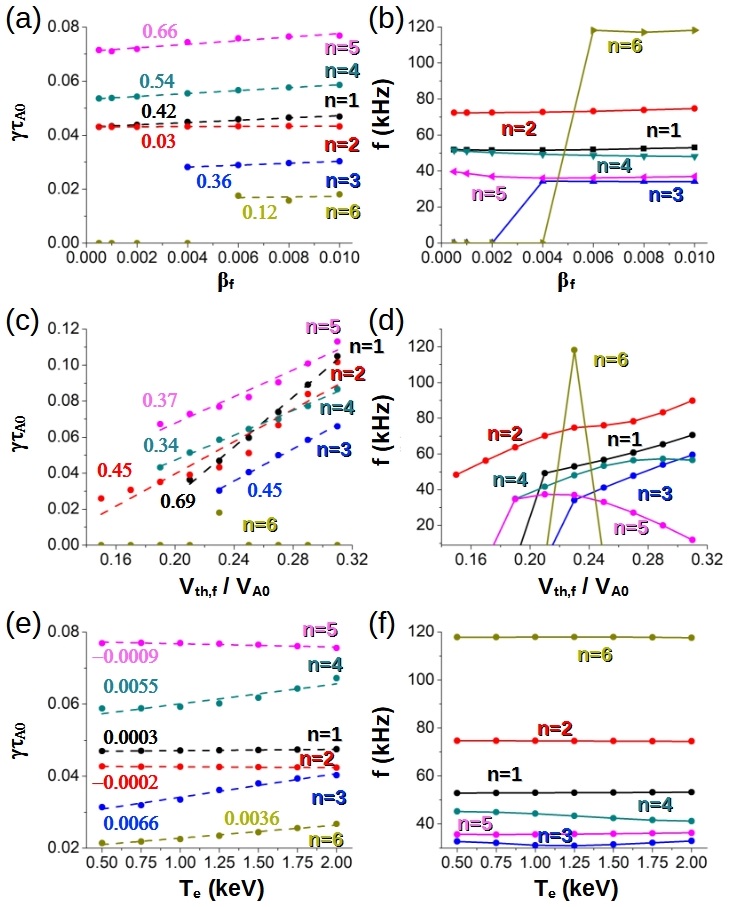}
\caption{AE/EPM stability trends of the sub-dominant modes in the MB configuration. (a) Growth rate and (b) frequency of the $n=1$ to $6$ modes for different EP $\beta$ values. (c) Growth rate and (d) frequency of the $n=1$ to $6$ modes for different $v_{th,f0} / V_{A0}$ ratios. (e) Growth rate and (f) frequency of the $n=1$ to $6$ modes for different electron temperatures. The dashed lines indicate the result of the linear regressions: $\gamma \tau_{A0} = A\beta_{f}$, $\gamma \tau_{A0} = B(v_{th,f0} / V_{A0})$ and $\gamma \tau_{A0} = CT_{e}$.}\label{FIG:9}
\end{figure}

Due to the variety of the alfvenic activity observed in the experiments and FAR3d simulations, the AE/EPM stability trends for dominant and sub-dominant modes are analyzed separately. Figure~\ref{FIG:8} is dedicated to the dominant modes and the figure~\ref{FIG:9} to the sub-dominant modes. 

Figure~\ref{FIG:8} panel a and b show the EP $\beta$ threshold is $0.001$ for the $n=3$ mode, $0.002$ for $n=1,2,4,5$ modes and $0.006$ for the $n=6$ mode. The linear regression of the growth rate indicates a slope of $4.85$ for the $n=1$ GAE, up to $19.7$ for the $n=6$ TAE, showing stronger trends compared to the $n=1$ and $2$ modes in the LB configuration. Figure~\ref{FIG:8} panels c and d indicate an increment of the $n=1$ to $3$ modes growth rate and frequency as the $v_{th,f0} / V_{A0}$ ratio increases, although the growth rate of the $n=4$ to $6$ modes drops if the ratio $v_{th,f0} / V_{A0} \geq 0.23$. That means, an increase of the EP slowing down time leads to a destabilizing effect on $n \leq 3$ modes although stabilizing on $n > 3$ modes. The data regressions show similar slopes for the $n=1$ and $2$ with respect to the LB configurations. On the other hand, if $v_{th,f0} / V_{A0} \geq 0.23$, the regressions slope for the $n=4$ to $6$ modes is negative, $3$ to $4$ times larger with respect to the regression slope for $v_{th,f0} / V_{A0} < 0.23$, pointing out a strong stabilizing effect as the EP energy increases. Figure~\ref{FIG:8} panels e and f show a weak decrease of $n=1$ to $6$ modes growth rate as the thermal $\beta$ increases. Nevertheless, the combined effect of the continuum, FLR and e-i Landau dampings as well the drop of the plasma resistivity as the electron temperature increases, have a weak stabilizing effect on the $n=1$ to $4$ modes, although slightly larger for the $n=5$ and $6$ modes. The data regression show a slope $2$ times smaller with respect to the LB configuration. Consequently, the stability trends suggest $n \leq 3$ modes cannot be stabilized by an increment of the ECH injection power, at least for the range of parameters evaluated in the present study. On the other hand, $n > 3$ modes can be stabilized by an increment of the ECH injection power, because a larger EP energy and thermal $\beta$ have a stabilizing effect. 

Figure~\ref{FIG:9} panels a and b indicate the EP $\beta$ threshold of $n=1,2,4,5$ modes is $0.0005$, $0.004$ for $n=3$ and $0.06$ for $n=6$. The data regression shows slopes between $0.03 – 0.66$, one order of magnitude smaller compared to the dominant modes in LB and MB configurations. Figure~\ref{FIG:9} panels c and d indicate an increase of the growth rate of the $n=1$ to $5$ modes with $v_{th,f0} / V_{A0}$ ratio. The data regression shows larger slopes compared to the dominant modes analysis. It should be noted that the mode $n=6$ is only unstable if $v_{th,f0} / V_{A0} = 0.23$. Figure~\ref{FIG:9} panels e and f indicate a small decrease of the growth rate for the $n=2$ and $5$ modes as the plasma thermal $\beta$ increases, although increasing for the other modes. This result can be explained by the type of sub-dominant modes destabilized, mainly Beta induced AEs (BAE). BAEs are triggered due to the coupling of an Alfven shear wave and a sound wave, thus an increase of the plasma temperature causes more energetic sound waves and a stronger resonance. The data regression indicates small negative slopes for the $n=2$ and $5$ modes, a small positive slope for the $n=1$ although larger positive slopes for the rest of the modes. Consequently, low frequency modes are weakly affect or further destabilized as $T_{e}$ increases, consistent with the observation of unstable modes with frequencies below $75$ kHz in the experiments.

\begin{figure}[h!]
\centering
\includegraphics[width=0.45\textwidth]{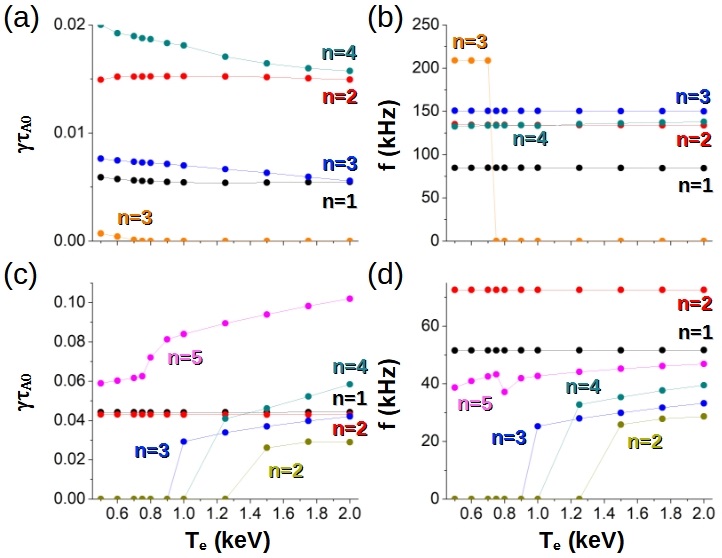}
\caption{High frequency modes (a) growth rate and (b) frequency in MB discharges as the electron temperature increases. Low frequency modes (c) growth rate and (d) frequency in MB discharges as the electron temperature increases. EP $\beta = 0.003$.}\label{FIG:10}
\end{figure}

Figure~\ref{FIG:10} shows the growth rate and frequency of high frequency and low frequency AE/EPM for model parameters reproducing the experimental conditions. Panel a and b indicate the modes in the frequency range between $80$ to $130$ kHz are not stabilized as the electron temperature increases from $0.5$ to $1$ keV, because the growth rate remains almost unchanged. On the other hand, there is a weak stabilizing effect on the mode around $150$ kHz as $T_{e}$ increases. On the other hand, the mode around $200$ kHz is fully stabilized. Panels c and d show several modes with frequencies between $40$ to $70$ kHz that remain unstable as $T_{e}$ increases. In addition, low frequency AEs (BAEs) are destabilized if $T_{e} \geq 1$ keV. The AE/EPM stability trends calculated in the simulations are consistent with the experimental observations.

Summarizing, the ECH injection in MB configurations causes the partial or full stabilization of modes with frequencies above $150$ kHz, although modes in the frequency range between $80$ to $150$ kHz are weakly affected. The stabilizing effect on high frequency modes is linked to the increment of the EP slowing down time, leading to a weakening of the resonance, combined with enhanced FLR, e-i Landau and continuum damping effects. On the other hand, low frequency modes with $f < 80$ kHz are weakly affect or further destabilized as $T_{e}$ increases, because the energy of the sound wave is higher leading to the destabilization of BAEs.

\subsection{High bumpiness configuration}

\begin{figure}[h!]
\centering
\includegraphics[width=0.45\textwidth]{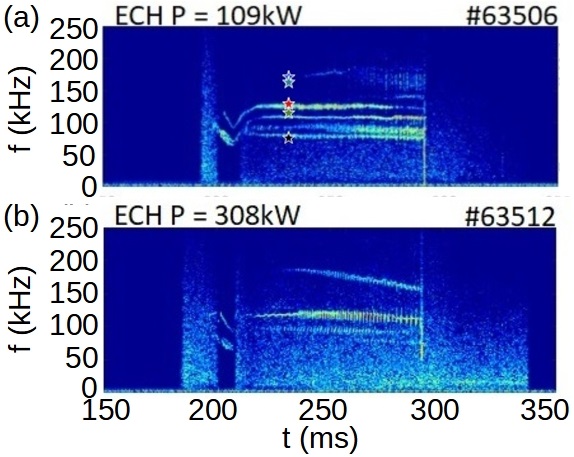}
\caption{Magnetic spectrogram of discharges with an ECH injection power of (a) $100$ kW and (b) $300$ kW. The colored stars indicate the frequency range of the dominant and sub-dominant modes calculated by FAR3d (black $n=1$, red $n=2$, $n=3$ blue, $n=4$ cyan and $n=1-5$ green).}\label{FIG:11}
\end{figure}

Figure~\ref{FIG:11} shows the magnetic spectrogram of two discharges with different ECH injection power for a high bumpiness configuration. Increase the ECH injection power leads to the stabilization of modes with a frequency above $125$ kHz, except for a high frequency mode with $f \approx 175$ kHz that is further destabilized. On the other hand, several modes with frequencies between $75$ and $125$ kHz are partially stabilized. The colored stars show the frequency of the dominant modes calculated by FAR3d for an EP $\beta = 0.003$, $T_{f} = 14$ keV and $T_{e} = 0.5$ keV (eigenfunctions are shown in the Appendix 1). The fastest growing modes are the $1/2$ GAE with $75$ kHz, $2/4$ GAE with $130$ kHz, $3/5$ GAE with $173$ kHz, $4/7$ GAE with $167$ kHz and $1/2-5/9$ helical AE (HAE) with $124$ kHz.

\begin{figure}[h!]
\centering
\includegraphics[width=0.45\textwidth]{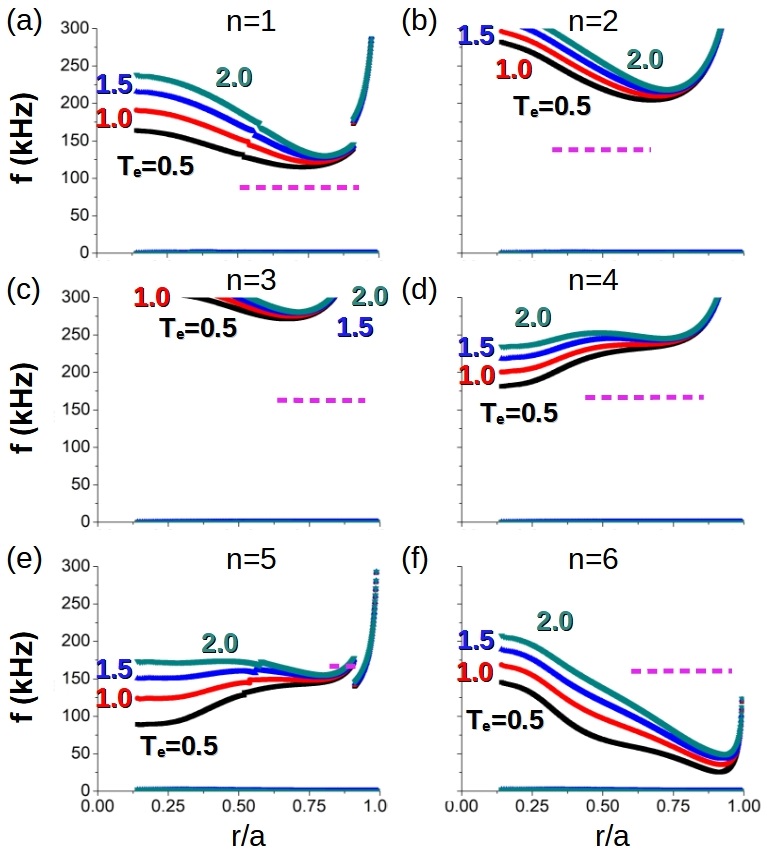}
\caption{Alfv\'en gap structure for different electron temperatures in the HB configuration for the (a) $n=1$, (b) $n=2$, (c) $n=3$, (d) $n=4$, (e) $n=5$ and (f) $n=6$ toroidal mode families. $T_{e} = 0.5$ keV (black line), $T_{e} = 1.0$ keV (red line), $T_{e} = 1.5$ keV (blue line), $T_{e} = 2.0$ keV (cyan line). The dashed pink horizontal lines indicate the frequency range and eigenfunction width of the dominant AE/EPM if EP $\beta = 0.01$.}\label{FIG:12}
\end{figure}

Figure~\ref{FIG:12} indicates the evolution of the Alfven gaps as the electron temperature increases from $0.5$ to $2$ keV. As observed in the other Heliotron J configurations, a larger electron temperature causes a frequency up-shift of the Alfv\'en gaps in the inner-middle plasma region. It should be noted that there is a helical gap linked to the helical family $n=1$\&$5$ at the plasma periphery in the frequency range around $150$ kHz. The horizontal pink dashed lines indicate the frequency range and radial location of the dominant AE/EPM obtained in FAR3d simulations with $T_{e} = 1$ keV and EP $\beta = 0.01$. The unstable modes are located in the middle - outer plasma region with frequencies from $80$ to $170$ kHz.

\begin{figure}[h!]
\centering
\includegraphics[width=0.45\textwidth]{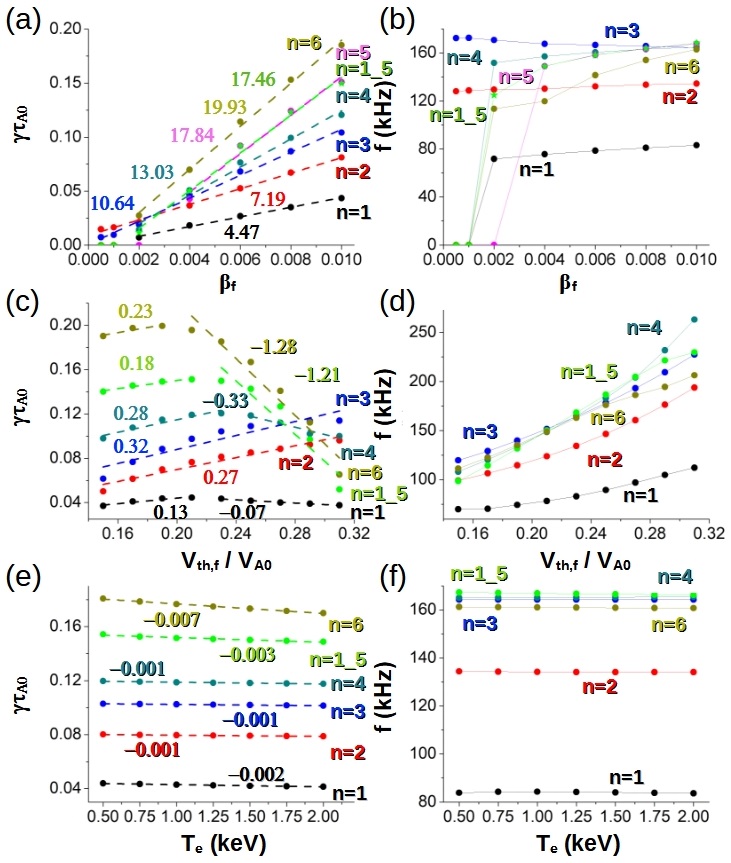}
\caption{AE/EPM stability trends of the dominant modes in the HB configuration. (a) Growth rate and (b) frequency of the $n=1$ to $6$ modes for different EP $\beta$ values. (c) Growth rate and (d) frequency of the $n=1$ to $6$ modes for different $v_{th,f0} / V_{A0}$ ratios. (e) Growth rate and (f) frequency of the $n=1$ to $6$ modes for different electron temperatures. The dashed lines indicate the result of the linear regressions: $\gamma \tau_{A0} = A\beta_{f}$, $\gamma \tau_{A0} = B(v_{th,f0} / V_{A0})$ and $\gamma \tau_{A0} = CT_{e}$.}\label{FIG:13}
\end{figure}

Figure~\ref{FIG:13} panel a and b indicate the EP $\beta$ threshold is $0.002$ for the $n=1,4,5,6$ toroidal mode families as well as for the helical family $n=1$\&$5$, $0.0005$ for the $n=2$ and $3$ modes. The mode $n=5$ and the helical family $n=1$\&$5$ have a similar growth rate and frequency, pointing out the effect of the helical coupling is rather weak. In the following, only the stability trends with respect to the helical family $n=1$\&$5$ are analyzed. The slope of the linear regression is $4.47$ for the $n=1$ GAE increasing to $19.9$ for the $n=6$ TAE, similar trends compared to the MB configuration although larger compared to the LB case. Figure~\ref{FIG:13} panels c and d show the decrease of the growth rate if $v_{th,f0} / V_{A0} \geq 0.23$ for all the modes except $n=2$ and $n=3$. Thus, for the $n=1,4,6$ modes and the helical family $n=1$\&$5$ there is an stabilizing effect linked to the increment of the EP slowing down time. The linear regression of the $v_{th,f0} / V_{A0}$ ratio indicates a negative slope $5$ times larger compared to the positive slope if $v_{th,f0} / V_{A0} > 0.23$ for the $n=6$ and $n=1$\&$5$ helical families. The negative slope is similar or slightly smaller with respect to the positive slope if $v_{th,f0} / V_{A0} < 0.23$ for $n=4$ and $n=1$ modes, respectively. Figure~\ref{FIG:13} panels e and f indicate a decrease of the growth rate for all the modes analyzed. Consequently, the increment of the electron temperature causes an enhancement of the continuum, FLR and e-i Landau dampings, leading to a stabilizing effect on all the modes. The data regression shows a rather weak stabilizing effect for the $n=2$ to $4$ modes (smaller compared to LB configuration), although larger for the $n=1$, $n=1$\&$5$ and $n=6$ modes.

\begin{figure}[h!]
\centering
\includegraphics[width=0.45\textwidth]{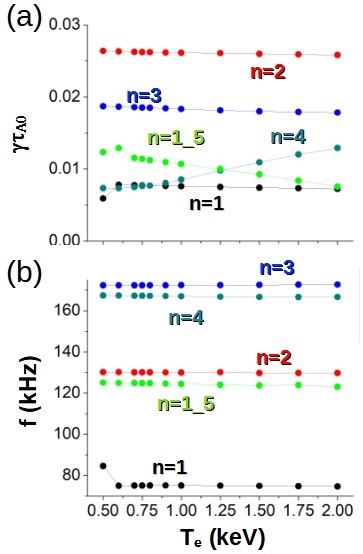}
\caption{Dominant modes (a) qrowth rate and (b) frequency in HB discharges as the electron temperature increases. EP $\beta = 0.003$.}\label{FIG:14}
\end{figure}

Figure~\ref{FIG:14} shows the growth rate and frequency of the dominant AE/EPM if the model parameters are similar to the experimental conditions as the electron temperature increases. The simulations indicate a weak stabilizing effect of the ECH injection on the $2/4$ GAE with $130$ kHz and the $1/2$ GAE with $f = 75$ kHz, although larger for the $1/2-5/9$ HAE with $125$ kHz. On the other hand, the $4/7$ GAE with $165$ kHz is further destabilized. In summary, the simulation results show a reasonable consistency with the experimental data.

In summary, the simulations show a weak stabilizing effect of the ECH injection in the HB configuration. The exceptions are the partial stabilization of the $1/2-5/9$ HAE with $125$ kHz and the further destabilization of the $4/7$ GAE with $f = 165$ kHz. The $1/2-5/9$ HAE is stabilized due to the increment of the EP slowing down time and the damping effects as $T_{e}$ increases. On the other hand, the $4/7$ GAE is further destabilized because the damping effects cannot compensate the destabilization induced by a larger EP slowing down time, stabilizing only above $v_{th,f0} / V_{A0} = 0.25$.

\section{Conclusions and discussion \label{sec:conclusions}}

The effect of the ECH injection power on the AE/EPM stability is analyzed for different magnetic configurations of Heliotron J. The stability trends linked to the variation of the EP $\beta$, thermal plasma $\beta$ and EP slowing down time with respect to the thermal electron temperature are identified. The simulations performed by the FAR3d code show a reasonable agreement with the experimental data, reproducing the effect of the ECH on the AE/EPM stability for different injection powers.

The simulations indicate the ECH injection modifies of the Alfven gap structure, that is to say, the continuum damping effect changes. There is a frequency up-shift of the gaps in the inner-middle plasma region and an outwards drift of the gap frequency minima, leading to a more localized and slender eigenfunction of the AE/EPM. In addition, a smaller eigenfunction width causes an enhanced of the thermal and EP FLR dampings, because the eigenfunction width is closer to the Larmor radius of thermal ions and EP, further reducing the instability free energy. On top of that, the electron-ion Landau damping increases and the plasma resistivity decreases as the plasma temperature grows, both stabilizing trends for the AE/EPM. On the other hand, a higher thermal plasma temperature also leads to a larger EP slowing down time, effect that can be stabilizing or destabilizing depending on the perturbation characteristics, for example the instability mode number, radial location or alfvenic family, between others. Likewise, a higher plasma temperature causes an increment of the EP $\beta$ and the further destabilization of the AE/EPM.

The analysis reproduces the AE/EPM stabilization in low bumpiness configuration as the ECH power increases. The modes are stabilized due to the enhancement of the continuum, FLR and electron-ion Landau dampings as the thermal $\beta$ increases. That means, the further destabilization caused by a larger EP $\beta$ and slowing down time is compensated by the stabilizing effect of the dampings.

The simulations find a diversity of unstable AE/EPM in medium bumpiness configuration. Furthermore, the stability of modes at some frequency ranges is almost independent of the ECH injection power, behavior explained by the counter balance between different stability trends. The increase of the plasma temperature causes the partial or full stabilization of modes with frequencies above $150$ kHz, because the increment of the EP slowing down time and the enhancement of the dampings have a stabilizing effect. On the other hand, some modes with frequencies between $80$ and $150$ kHz are not stabilized, because a larger EP slowing down time causes the modes further destabilization, not compensated by the damping effects. Likewise, modes with a frequency below $80$ kHz can be further destabilized as the plasma temperature increases. This is the case of modes belonging to the Beta induced AE family, triggered due to the resonance of Alfven and sound waves. A higher plasma temperature boost the sound waves energy leading to the resonance enhancement, reason why the simulations show an increment of the modes growth rate. Nevertheless, the growth rate of the BAEs is $2$ to $3$ times smaller compared to the dominant modes at a higher frequency.

The numerical study shows a smaller AE/EPM activity in high bumpiness configuration compared to the medium bumpiness case. Besides that, the increment of the ECH injection power does not lead to the AE/EPM stabilization. The simulations indicate weak trends between plasma temperature and AE/EPM stability. The exceptions are the stabilizing effect calculated for the $1/2-5/9$ HAE with $125$ kHz and the destabilizing effect found for the $4/7$ GAE with $f = 165$ kHz. The decrease of the $1/2-5/9$ HAE growth rate is caused by the combined effect of the dampings enhancement and the resonance weakening as the EP slowing down time increases. On the other hand, the growth rate of the $4/7$ GAE increases because the dampings are not large enough to compensate the resonance enhancement as the EP slowing down time increases.

It must be mentioned that the effect of the ECH injection on the EP pitch angle scattering is not included in the model, only passing EP are considered in the study. Consequently, the analysis of the EP pitch angle scattering and the consequences in the AE stability require a more sophisticated numerical model.

The present study reveals the complex interconnection of the different AE/EPM stability trends as the plasma parameters change with the ECH injection power. Such interdependence explains the variety of experimental results obtained in the different Heliotron J magnetic configurations as well as other devices if the ECH is applied. An increase of the thermal $\beta$ as the electron temperatures grows leads to an enhancement of the continuum, FLR and e-i Landau dampings, stabilizing effect that can be counter-balanced by the increment of the EP slowing down time and EP $\beta$ depending on the AE/EPM properties. In summary, ECH is a promising tool to improve the AE/EPM stability in nuclear fusion devices, although its effect could be weak or even harmful depending on the unstable mode characteristics, particularly if the modes are BAEs. In addition, the ECH injection is not recommendable in plasma regions where AEs of different Alfvenic families and toroidal mode numbers overlap, leading to the full or partial stabilization of some modes although an almost null or even further destabilization of others. On the other hand, discharges with a reduced number of AE/EPM enable a more efficient application of the ECH, that is to say, an optimal identification of the ECH injection power and deposition region can lead to the mode stabilization. Efficient examples of ECH injection are, for example, low bumpiness discharges in Heliotron J stabilizing the $n=1$ EPM and $n=2$ GAE \cite{19} as well as LHD discharges with stable EIC \cite{20}. 

Present results should be consider as the first step of a global research line dedicated to analyze the ECH effect on the AE stability. Future Heliotron J experiments will be dedicated to isolate the different effects of the ECH on the plasma stability. In addition, the target of coming studies will be the AE stability of LHD and TJ-II plasma heated by NBI and ECH. In the mid-term, the analysis will be extended to the case of tokamak devices as DIII-D, JET and ASDEX. These studies will provide further validation of the present modeling results, particularly the stability trends identified, providing new tools to improve the plasma heating efficiency of future nuclear fusion devices.

\section*{Appendix}

\subsection*{Analysis of the instabilities in MB configuration}

\begin{figure}[h!]
\centering
\includegraphics[width=0.45\textwidth]{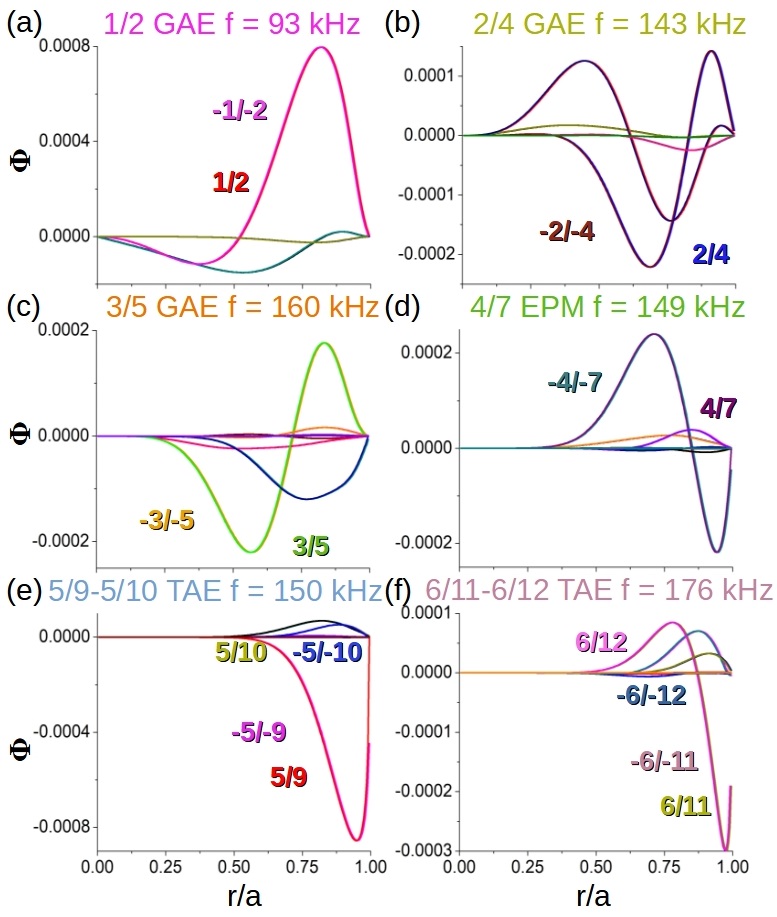}
\caption{Eigenfunction of the dominant modes calculated by FAR3d in the MB configuration: (a) $1/2$ GAE, (b) $2/4$ GAE, (c) $3/5$ GAE, (d) $4/7$ EPM, $5/9-5/10$ TAE and $6/11-6/12$ TAE.}\label{FIG:X1}
\end{figure}

\begin{figure}[h!]
\centering
\includegraphics[width=0.45\textwidth]{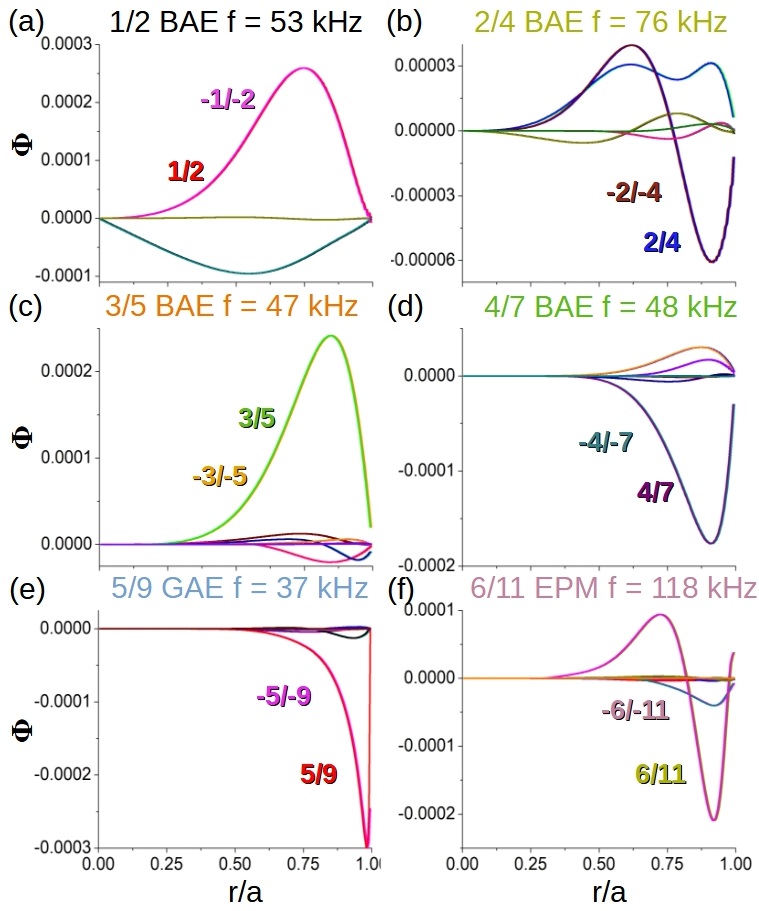}
\caption{Eigenfunction of the sub-dominant modes calculated by FAR3d in the MB configuration: (a) $1/2$ BAE, (b) $2/4$ BAE, (c) $3/5$ BAE, (d) $4/7$ BAE, $5/9$ GAE and $6/11$ EPM.}\label{FIG:X2}
\end{figure}

Figures~\ref{FIG:X1} and~\ref{FIG:X2} show the eigenfunction of the dominant and sub-dominant modes calculated by the FAR3d code if the EP $\beta = 0.01$, $T_{f} = 14$ keV and $T_{e} = 1$ keV in the MB configuration. The eigenfunctions are plotted for an EP $\beta$ three times higher compared to the experiment, showing a more robust instability to easily identify the perturbation properties. The dominant modes are the $1/2$ GAE with $f=93$ kHz, the $2/4$ GAE with $f = 143$ kHz, the $3/5$ GAE with $160$ kHz, the $4/7$ EPM with $f = 149$ kHz, the $5/9-5/10$ TAE with $f = 150$ kHz and $6/11-6/12$ TAE with $f = 176$ kHz. Below $100$ kHz, the sub-dominant modes showing the largest growth rates are mainly BAEs with $f=37$ to $76$ kHz, except the $6/11$ EPM with $f = 118$ kHz. All the modes are unstable in the middle-outer plasma region, showing a slender eigenfunction further localized at the plasma periphery as the toroidal mode number increases.

\subsection*{Analysis of the instabilities in HB configuration}

\begin{figure}[h!]
\centering
\includegraphics[width=0.45\textwidth]{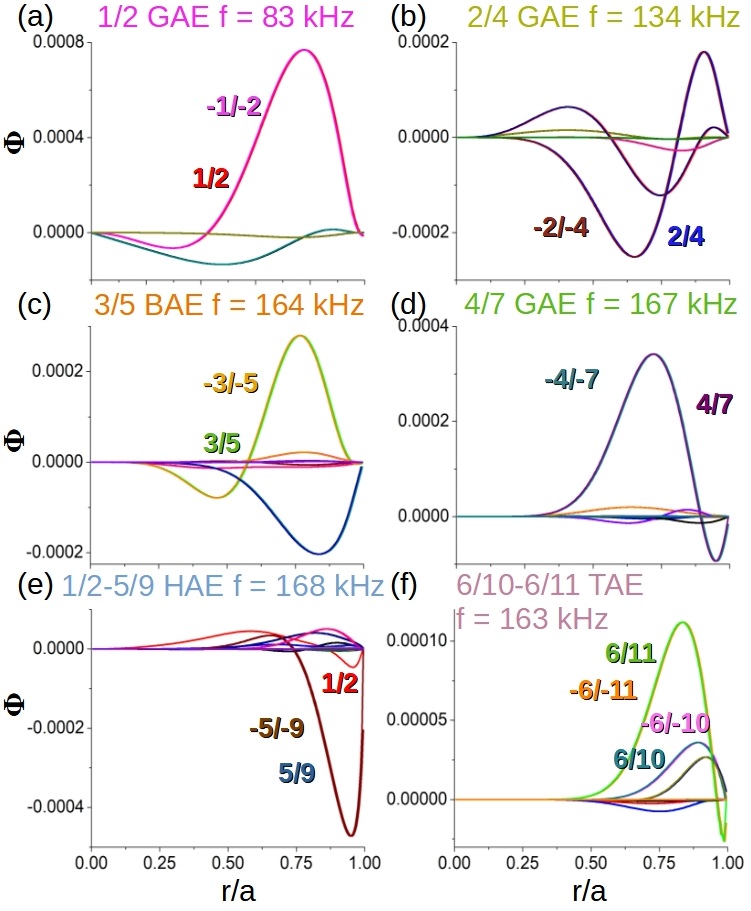}
\caption{Eigenfunction of the dominant modes calculated by FAR3d in the HB configuration: (a) $1/2$ GAE, (b) $2/4$ GAE, (c) $3/5$ BAE, (d) $4/7$ GAE, $1/2-5/9$ HAE and $6/10-6/11$ TAE.}\label{FIG:X3}
\end{figure}

Figures~\ref{FIG:X3} shows the eigenfunction of the dominant modes if the EP $\beta = 0.01$, $T_{f} = 14$ keV and $T_{e} = 1$ keV in the HB configuration. The $1/2$ GAE with $f = 83$ kHz and the $2/4$ GAE with $f = 134$ kHz are unstable. At higher frequencies, in the range of $160$ to $170$ kHz the $3/5$ BAE, $4/7$ GAE, $1/2-5/9$ HAE and $6/10-6/11$ TAE are triggered. All the modes are destabilized in the middle-outer plasma region and a larger toroidal mode number leads to a more localized and slender eigenfunction at the plasma periphery.

\ack

The authors would like to thank the Heliotron J technical staff for their contributions in the operation and maintenance of Heliotron J. This work was supported by the Comunidad de Madrid under the project 2019-T1/AMB-13648, Comunidad de Madrid - multiannual agreement with UC3M (“Excelencia para el Profesorado Universitario” - EPUC3M14 ) - Fifth regional research plan 2016-2020 as well as NIFS Collaborative Research Program NIFS08KAOR010, NFIS10KUHL030, NIFS07KLPH004 and "PLADyS" JSPS Core-to-Core Program, A. Advanced Research Networks. 

\hfill \break

\end{document}